\documentclass{ifacconf}

\usepackage{graphicx}      
\usepackage{amsmath,amssymb}
\usepackage{mathtools} 
\usepackage{wrapfig}
\usepackage{dblfloatfix}    
\usepackage{fancyhdr} 

\renewcommand{\Re}{\operatorname{\mathbb{R}e}}
\renewcommand{\Im}{\operatorname{\mathbb{I}m}}

\makeatletter
\let\old@ssect\@ssect 
\makeatother

\usepackage{natbib}
\usepackage{hyperref}

\makeatletter
\def\@ssect#1#2#3#4#5#6{%
  \NR@gettitle{#6}
  \old@ssect{#1}{#2}{#3}{#4}{#5}{#6}
}
\makeatother

\providecommand{\url}[1]{\texttt{#1}}

\newcommand{\doi}[1]{\href{https://doi.org/#1}{doi:#1}}

\begin{document}
\begin{frontmatter}

\title{Force-Limited Control of Wave Energy Converters using a Describing Function Linearization\thanksref{footnoteinfo}} 

\thanks[footnoteinfo]{This material is based on work supported by National Science Foundation Graduate Research Fellowship Grant No. DGE–2139899.}

\author[First]{Rebecca McCabe} 
\author[Second]{Maha N. Haji} 
\address*{Sibley School of Mechanical and Aerospace Engineering, \\Cornell University, 
   Ithaca, NY 14853 USA }
   \address[First]{e-mail: rgm222@cornell.edu}
   \address[Second]{e-mail: maha@cornell.edu}

\begin{abstract}                
Actuator saturation is a common nonlinearity. In wave energy conversion, force saturation conveniently limits drivetrain size and cost with minimal impact on energy generation. However, such nonlinear dynamics typically demand numerical simulation, which increases computational cost and diminishes intuition. This paper instead uses describing functions to approximate a force saturation nonlinearity as a linear impedance mismatch. In the frequency domain, the impact of controller impedance mismatch (such as force limit, finite bandwidth, or parameter error) on electrical power production is shown analytically and graphically for a generic nondimensionalized single degree of freedom wave energy converter in regular waves. Results are visualized with Smith charts. Notably, systems with a specific ratio of reactive to real mechanical impedance are least sensitive to force limits, a criteria which conflicts with resonance and bandwidth considerations. The describing function method shows promise to enable future studies such as large-scale design optimization and co-design.
\end{abstract}


\end{frontmatter}


\section{Introduction}
\vspace{-10pt}
\subsection{Motivation}
Ocean wave energy converters (WECs) are an immature yet promising source of renewable energy to decarbonize coastal grids and offshore systems. Maximum power transfer requires impedance matching of WEC controls, powertrain, and hydrodynamics, but plant uncertainty, controller bandwidth, and physical constraints prevent perfect matching. Actuator force limits are especially relevant given waves' high-force, low-speed nature and the scaling of device cost with force. WEC controllers must maximize power while obeying force limits, although the resulting nonlinearity requires computationally-costly numerical optimization. There is a gap of rapid and intuitive methods suitable for early tradeoff analysis. To this end, the present paper demonstrates linear analytic treatment of force saturation limits and other sources of impedance mismatch.

\subsection{Literature Review}
Prior studies on WEC powertrain constraints investigate position, velocity, force, rate of change of force, and power flow direction. Constrained numerical optimization is typically used. For example, \cite{faedo_optimal_2017} review model predictive control while \cite{strofer_control_2023} apply pseudo-spectral methods.

A minority of work tackles the problem analytically. \cite{zou_optimal_2017} use the Pontryagin principle to show that a bang-singular arc-bang controller is optimal, revealing that saturating the unsaturated optimal solution can still be optimal in certain cases. 
They derive analytical piecewise expressions for the optimal control force, but finding the corresponding power still requires numerical simulation.
\cite{bacelli_geometric_2013} present a geometric tool to analyze simultaneous force and position constraints. They provide analytical relationships between power and root-mean-square signals, which relate to upper and lower bounds on constraint values, but the bounds are not tight. \cite{merigaud_geometrical_2023} introduce another geometric tool which accounts only for position constraints and focuses on hydrodynamics over powertrain. The tool is visually and mathematically similar to a Smith chart. To the authors' knowledge, no prior work uses describing functions to address WEC constraints, though \cite{flower_describing-function_1980} use them to model a WEC with nonlinear damping, and \cite{quartier_influence_2021} use the same technique to model drag, calling it a Fourier approximation. Outside of wave energy, \cite{fukui_impedance_2021} apply describing functions to model velocity and torque saturation on an impedance-controlled robot and experimentally validate the results.



\subsection{Paper Outline and Contribution}
Section~\ref{sec:linear} presents the paper's first contribution: application of linear theory to analyze the relationship between impedance mismatch, signal amplitude, and power for a WEC in regular waves. Explicit analytical expressions are derived, and results are visualized on a Smith chart. As the second contribution, section~\ref{sec:nonlinear} suggests that constraints typically solved numerically, such as force limits, be linearized with describing functions in order to apply the previous section's results. The approximation is derived and discussed in the context of impedance mismatch. 

These contributions provide three main benefits. First, they offer an intuitive process for rapid tradeoff analysis early in the design process, with the opportunity to apply standard linear frequency-domain tools. Second, they are computationally efficient enough to integrate with more expensive techniques like design optimization or control co-design, either directly or as an initial guess for solvers of higher fidelity dynamics. Third, the results are analytical and differentiable, providing gradients for sensitivities or to accelerate convergence of outer optimizations.

\section{Peak Limiting in the Linear Case}\label{sec:linear}
\vspace{-8pt}\subsection{General Impedance-Mismatched System}\label{sec:generic}
\begin{figure}[htbp!]
\begin{center}
\includegraphics[width=0.4\linewidth]{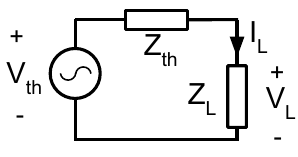}    
\vspace{-10pt}\caption{Thévenin equivalent circuit for linear system} 
\label{fig:circuit}
\end{center}
\end{figure}

The analysis starts with a generic linear system modeled as a Thévenin equivalent circuit with AC voltage source $V_{th}$ and complex source impedance $Z_{th}$, shown in Fig.~\ref{fig:circuit}.

The load impedance $Z_L$ is to be selected, with the conflicting goals of maximizing average power transfer $\overline{P}_L$ and minimizing the peak amplitude of load current $|I_L|$ or voltage $|V_L|$. Maximum power transfer occurs when there is impedance matching, meaning $Z_L = Z_{th}^*$ where $*$ indicates complex conjugate. The load average power, peak voltage, and peak current at this matched point, denoted $\overline{P}_L^m, |V_L^m|,$ and $|I_L^m|$ respectively, are found as:
\begin{equation}\label{eq:matched-values}
\hspace{-.55em}
    \overline{P}_L^m = \frac{|V_{th}|^2}{8 \Re(Z_{th})}, 
    |V_L^m| = \frac{|V_{th}| |Z_{th}|} {2 \Re(Z_{th})}, 
    |I_L^m| = \frac{|V_{th}|}{2 \Re(Z_{th})}
\end{equation}
where $\Re$ means the real part and $\Im$ the imaginary part.

To consider all possibilities of the unmatched case, we set $Z_L = z Z_{th}^*$ for arbitrary complex number $z$. The space of $z$ can be visualized using a Smith chart, where $\Re(z)$ is on a curved horizontal axis and $\Im(z)$ is on a curved vertical axis. The axes are curved such that the chart can be simultaneously read as a standard polar plot of the complex reflection coefficient $\Gamma$, which is a transformation of $z$ defined as $\Gamma = \frac{z-1}{z+1}$. The impedance-matched case of $z=1$, $\Gamma = 0$ is found at the center of the plot, the minimum voltage at $z=0$, $\Gamma = -1$ on the left, and the minimum current at $z \rightarrow \infty$, $\Gamma = 1$ on the right.

\begin{figure}[!thbp]
    \centering
    \includegraphics[width=\linewidth]{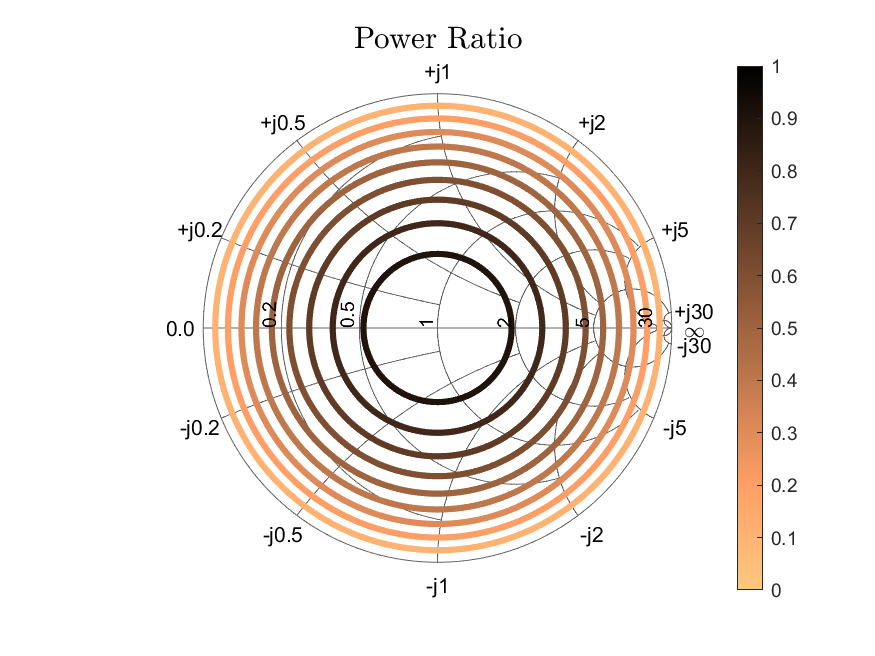}
   \vspace{-30pt}\caption{Smith chart showing average power, using (\ref{eq:ratio-power}).}
    \label{fig:power-smith}
\end{figure}

The average power, peak voltage, and peak current in the unmatched case can be found using standard circuit techniques and expressed as fractions of their matched counterparts. \cite{roberts_conjugate-image_1946} derives the power ratio:
\begin{equation}\label{eq:ratio-power}
     \frac{\overline{P}_L}{\overline{P}_L^m} = 1 - |\Gamma|^2
\end{equation}
This relationship is visualized on the Smith chart in Fig.~\ref{fig:power-smith}. As the impedance ratio $z$ gets further away from the impedance-matched condition $z=1$ at the center of the circle, the power lowers quadratically.
\begin{figure*}[tb!]
    \centering
    \includegraphics[width=18.2cm]{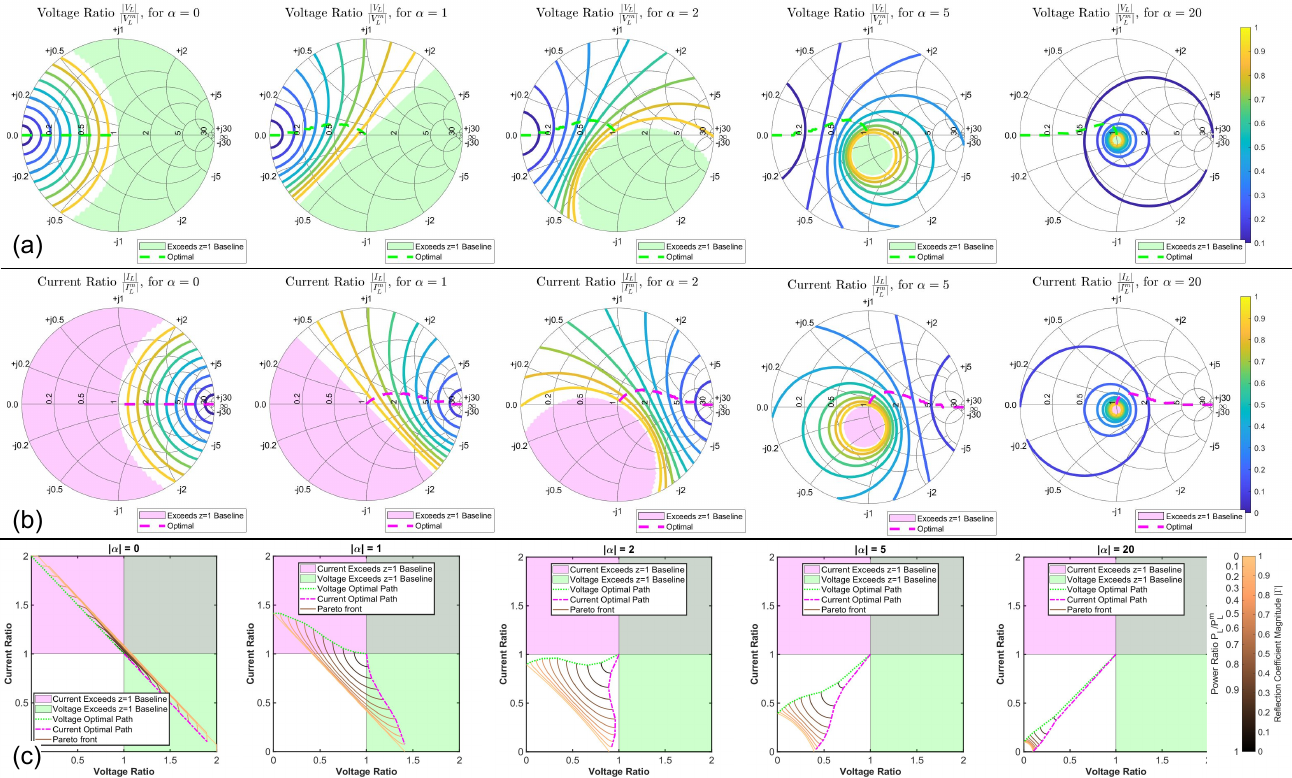}
    \caption{Smith charts showing the (a) peak voltage and (b) peak current at the load for any $z=Z_L/Z_{th}^*$, paired with (c) the pareto tradeoff between voltage, current, and power. The Thévenin reactance parameter $\alpha$ is swept. Shaded regions indicate that the voltage (green) or current (pink) ratios exceed one. Dashed lines are optimal contours.}
    \label{fig:ratio-smith}
\end{figure*}

\vspace{-10pt}
The corresponding voltage and current ratios are:
\begin{equation}\label{eq:ratios}
    \frac{|V_L|}{|V_L^m|},\frac{|I_L|}{|I_L^m|} = \sqrt{\frac{|\Gamma|^2 +  2 \epsilon \Re(\Gamma) + 1}{\alpha^2 |\Gamma|^2 + 2 \alpha \Im(\Gamma) + 1} } 
\end{equation}
where $\epsilon$ indicates sign: $\epsilon=1$ for voltage ratio and $\epsilon=-1$ for current ratio, and $\alpha = \Im(Z_{th})/\Re(Z_{th})$ is a parameter related to the phase of the source impedance. These relationships are visualized on the Smith charts in Fig.~\ref{fig:ratio-smith} (a) and (b), where contours show the ratios as a function of $z$ (and thus $\Gamma$) for various values of $\alpha$. Only positive $\alpha$ (inductive $Z_{th}$) is shown for brevity. The contours for negative $\alpha$ (capacitive $Z_{th}$) can be found by reflecting the graphs over the horizontal axis ($\Im(\Gamma) = 0)$ due to symmetry. On the Smith charts, points where the voltage and current ratios exceed 1 are shaded. These points are undesirable because they absorb less power than the baseline $z=1$, $\Gamma=0$ matched case while having higher peaks. The optimal contours (lowest voltage and current ratios for a given power ratio) are traced out with dashed lines. Because the optimal voltage reduction path requires decreasing the impedance and the optimal current reduction path requires increasing the impedance, it is not possible to follow both paths simultaneously. For sufficiently high values of $|\alpha|$, it is possible to reduce both current and voltage simultaneously (i.e. avoid the shaded region of both Smith charts in Fig.~\ref{fig:ratio-smith} (a) and (b)), although for $\alpha=0$, decreasing voltage implies increasing current and vice versa. This tradeoff is explored further in section (c) of Fig.~\ref{fig:ratio-smith}, a pareto front showing the nondominated combinations of the voltage, current, and power ratios.

The optimal paths $(\cdot)^\textrm{opt}$ of Fig. \ref{fig:ratio-smith} are derived by setting the derivative of the voltage and current ratios (\ref{eq:ratios}) with respect to $\angle\, \Gamma$ equal to zero, for $\angle\, \Gamma= \tan^{-1} (\Im(\Gamma)/\Re(\Gamma))$:
\begin{equation}\label{eq:optimal-vi}
\hspace{-4pt} \angle\, \Gamma ^\textrm{opt} \hspace{-2pt} = 2 \tan^{-1}\hspace{-3pt}\left[ \frac{\alpha^2 |\Gamma|^2 + 1}{\sigma + \epsilon \alpha (1 + |\Gamma|)^2}\hspace{-2pt}\right] + \epsilon\cos^{-1} \hspace{-3pt}\left[ \frac{-2\alpha| \Gamma|}{\sigma} \right]
\end{equation}
where $\sigma = \sqrt{(\alpha^2|\Gamma|^2 + 1)^2 + \alpha^2 (|\Gamma|^2 + 1)^2}$ and sign indicator $\epsilon=1$ obtains the angle at minimum voltage and $\epsilon=-1$ minimum current. \cite{otoshi_maximum_1994} obtained a similar result for a two-port system. Further manipulation of (\ref{eq:optimal-vi}) shows that the minimum-voltage and minimum-current angles are supplementary: $\angle\, \Gamma^{\textrm{opt,V}}+ \angle\, \Gamma^{\textrm{opt,I}}= \pi$. Using this relation and plugging into (\ref{eq:ratios}) reveals that the ratios are equal at their respective optima: $\frac{|V_L|}{|V_L^m|}^\textrm{opt,V} \hspace{-8pt}= \frac{|I_L|}{|I_L^m|}^\textrm{opt,I}\hspace{-15pt}.\hspace{10pt}$ Intuitively, voltage and current reductions are symmetric.

\begin{figure}[hbt]
    \centering
    \includegraphics[width=.8\linewidth]{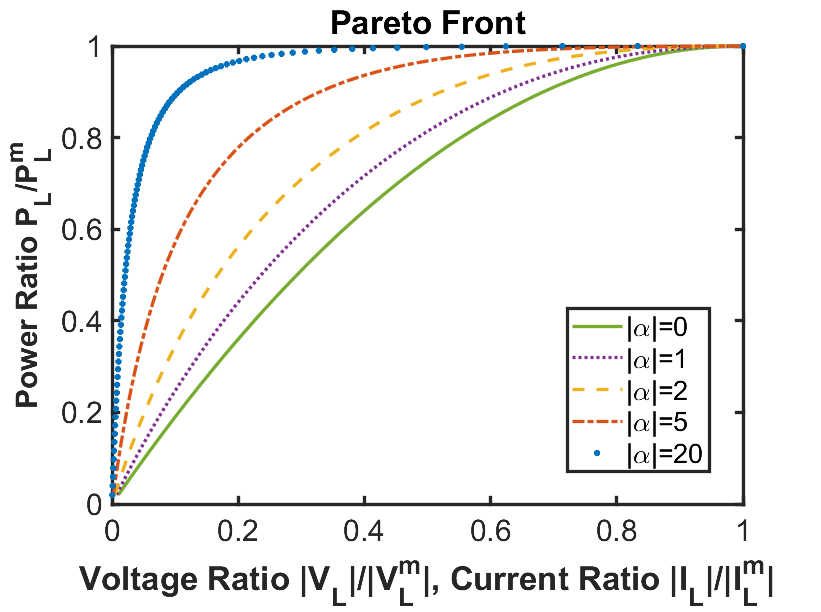}
  \vspace{-5pt}  \caption{Pareto front for optimal voltage and current ratios.}
    \label{fig:pareto}
\end{figure}

The optimal contours are aggregated in Fig.~\ref{fig:pareto} to show the tradeoff between power and voltage/current. Interestingly, as $|\alpha|$ grows (i.e., as $Z_{th}$ becomes less resistive and more reactive), there is less of a power penalty for a given voltage or current reduction. In other words, power in the pure reactive $Z_{th}$ case is least sensitive to voltage and current limits. For fixed $\Re(Z_{th})$,  (\ref{eq:matched-values}) shows the baseline matched power is independent of $\alpha$, suggesting that plant design should maximize $|\alpha|$.

Note that most of the optimal contours of Fig.~\ref{fig:ratio-smith} (a) and (b) require a $z$ with nonzero imaginary part, i.e. a $Z_L$ that is more or less reactive than the impedance-matched case, rather than merely scaled up or down. This highlights the difference between ``constrained optimal control" and ``optimal constrained control." The former refers to the unconstrained optimal controller that has been scaled or saturated until it meets the constraint, while the latter refers to the controller that is optimal for the constrained problem, which is distinct under the present assumption of linear control. Scaling down or saturating the control signal computed with the unconstrained optimal impedance yields a signal with a fundamental amplitude identical to one computed with a proportionally scaled linear control impedance. This enforces $\Im(z)=0$, which is evidently non-optimal for all but $\alpha=0$. This is distinct from the complex $z$ in the optimal profiles of (\ref{eq:optimal-vi}) and Fig.~\ref{fig:pareto}, which would be considered optimal constrained control. Interestingly, \cite{zou_optimal_2017} show that the optimal constrained and constrained optimal controllers are identical if nonlinear control is allowed. In summary, classical theory reveals the effect of impedance mismatch on power, current, and voltage, informing the choice of a load impedance to balance power generation and peak limiting constraints.

\subsection{Wave Energy Converters}
\begin{figure}
    \centering
    \includegraphics[width=\linewidth]{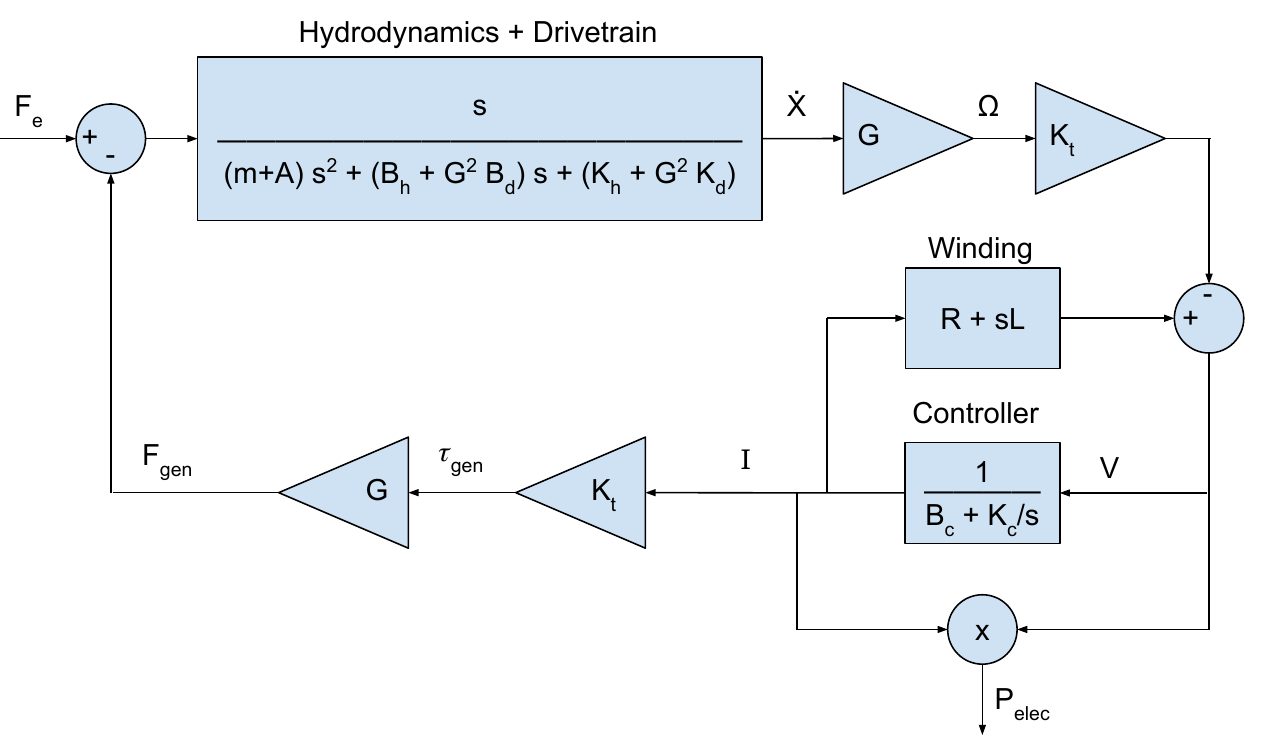}\vspace{-25pt}
    \caption{Block diagram of dynamics.}
    \label{fig:block-diagram-dynamics}
\end{figure}

\vspace{-5pt}
Applying the preceding analysis to a WEC requires a Thévenin equivalent circuit for WEC dynamics. This study assumes a single degree of freedom floating body coupled to a power take-off (PTO) with a drivetrain and a linear or rotational synchronous surface permanent magnet electric generator. Substituting a hydraulic or other impedance is straightforward provided the system remains linear. 
The generator model is non-ideal and the objective is electrical power. The frequency domain WEC dynamics are:
\begin{equation}\label{eq:wec-dynamics}
\begin{aligned}
    &((m+A)s^2 + B_h s + K_h) X + F_{P} = F_e &&\textrm{Body}\\
    &G \tau_{PTO} = F_{PTO},~G s X = \Omega &&\textrm{Gear ratio}\\
    &\tau_{PTO} = \left(B_d + K_d/s\right) \Omega + \tau_{gen} &&\textrm{PTO}\\
    &\tau_{gen} = K_t I,~V = I(R + s L) - K_t \Omega &&\textrm{Generator}\\
    &V = \left(B_c + K_c/s\right) I &&\textrm{Controller}\\
    &P_{elec} = 0.5 \Re(I^* V) &&\textrm{Power}
\end{aligned}
\end{equation}
with Laplace variable $s$, mass $m$, added mass $A$, hydrodynamic damping $B_h$, hydrostatic stiffness $K_h$, WEC position $X$, power take-off and wave excitation forces $F_{P}$ and $F_e$, effective gear ratio $G$, drivetrain mechanical stiffness and damping $K_d$ and $B_d$, generator torque and rotation speed $\tau_{gen}$ and $\Omega$, generator torque constant $K_t$, generator q-axis current and voltage $I$ and $V$, generator resistance and inductance $R$ and $L$, controller stiffness and damping $K_c$ and $B_c$, and average electrical power $P_{elec}$. The controller acts between $V$ and $I$ rather than $\tau_{gen}$ and $\Omega$ as is more typical, providing equivalent dynamics and making the relevant Thévenin equivalent more convenient to define. $B_d$ and $K_d$ can capture mooring and drag forces. Frequency dependence of hydrodynamic coefficients $A$, $B_h$, and $F_e$ is omitted since this study assumes regular waves. Fig.~\ref{fig:block-diagram-dynamics} shows a block diagram of the dynamics.

Since electrical, not mechanical, power is the focus, choose the controller as the load impedance: $Z_L = Z_c = B_c + \frac{K_c}{s}$. Thus the Thévenin equivalent load voltage/current are the generator q-axis voltage/current: $V_L = V, I_L = I$. Thévenin parameters from \cite{strofer_control_2023} are then:
\begin{equation}\label{eq:thevenin}
    Z_{th} = Z_w + \frac{K_t^2 G^2}{Z_m}, \quad  |V_{th}| = \frac{K_t G F_e}{|Z_m|}
\end{equation}
with winding impedance $Z_w = R + s L$ and mechanical impedance $Z_m = B_h + G^2B_d + (m+A)s + \frac{K_h + G^2 K_d}{s}$.
Substituting \eqref{eq:thevenin} into \eqref{eq:matched-values} and applying the Haskind relation between $B_h$ and $F_e$ yields the matched power:
\begin{equation}\label{eq:matched-power-wec}
     \overline{P}_L^m = \frac{G_0 J}{k}\frac{\mathcal{D}}{1+\frac{\mathcal{R}}{\mathcal{D}} (1+\alpha_m^2)}, ~G_0 = \begin{cases}
         1 & \mathrm{heave} \\
         2 & \mathrm{surge/pitch}
     \end{cases}
\end{equation}
with incident energy density $J$, wavenumber $k$, and gain $G_0$. \cite{zou_practical_2023} show this applies for any WEC shape. Nondimensional quantities $\mathcal{R}=R B_h/(K_t^2 G^2)$, $\mathcal{D}=B_h/(B_h+G^2B_d)$,  and $\alpha_m = \Im(Z_m)/\Re(Z_m)$ are the normalized resistance and ratios of hydrodynamic to total damping and reactive to real impedance, respectively.

This matched power is not possible when there is controller impedance mismatch, either intentional with the intent to obey a given amplitude limit, or unintentional due to parameter uncertainty or controller bandwidth limitations in broadband waves. The mismatched power $\overline{P}_L$ is found by multiplying the matched power $\overline{P}_L^m$  in \eqref{eq:matched-power-wec} by the power ratio in \eqref{eq:ratio-power}. The required reflection coefficient $\Gamma$ to meet constraints is found via (\ref{eq:ratios}) or graphically via Fig.~\ref{fig:ratio-smith} and \ref{fig:pareto}. Doing so requires an expression for $\alpha$, found via \eqref{eq:thevenin}:
\begin{equation}\label{eq:alpha}
    \alpha = \frac{\Im(Z_{th})}{\Re(Z_{th})} =  \frac{\mathcal{L}\mathcal{R} (1+\alpha_m^2) - \mathcal{D} \alpha_m}{\mathcal{R} (1+\alpha_m^2)+\mathcal{D}}
\end{equation}
with $\mathcal{L}=\frac{\omega L}{ R }$. Typically $\mathcal{L} \approx 0$ since the wave period $2 \pi / \omega$ far exceeds the winding electrical time constant $L/R$.

Section~\ref{sec:generic} allows choice of $\Gamma$ to limit q-axis variables $V$ and $I$, but limiting other quantities may be desired. Limiting $V$ has little direct utility. Instead, a limit on phase voltage $V_s$ represents the voltage of a vector drive which affects the generator's torque-speed curve; a limit on position $X$ represents actuator stroke or kinematic constraints; and a limit on apparent power $S$ sizes the PTO including reactive power. These amplitudes are:
\begin{equation}\label{eq:X-Vs}
\begin{aligned}
    X = \frac{F_e/s}{Z_m + \frac{K_t^2 G^2}{Z_w - z Z_{th}^*}}, \quad V_s^2 = V^2 + (L p \Omega I)^2 \\
\frac{S_{\max,\min}}{\overline{P}_L^m} = \frac{\overline{P}_L}{\overline{P}_L^m} \pm \frac{|V_L|}{|V_L^m|}\frac{|I_L|}{|I_L^m|}\sqrt{1+\alpha^2}
\end{aligned}
\end{equation}
where $p$ is the number of machine poles. Smith plots similar to Fig.~\ref{fig:ratio-smith} could be made to visualize the effect of these limits, but more parameters besides $\alpha$ must be swept to visualize the design space. Meanwhile, a generator force limit of $|F_{gen}| \leq F_{max}$ is achieved with a q-axis current limit of $|I| \leq I_{max} = \frac{F_{max}}{K_t G}$, which is well-captured by the earlier parameterization. For brevity, the rest of this paper focuses on current (force) limits, which drive PTO cost and are of primary design interest. Thus, $\Gamma$ is selected to set current ratio $\frac{|I_L|}{|I_L^m|} = \frac{I_{max}}{|I_L^m|}$, thereby enforcing the limit.

Besides facilitating control, the formulation also informs plant design. By inspection of \eqref{eq:matched-power-wec}, the maximum-power plant design in the matched (unconstrained) case is $(\mathcal{R,D},\alpha_m)^\mathrm{opt}=(0,1,0)$, intuitively minimizing loss due to resistance, friction, and reactive power. $\alpha_m$ relates to the uncontrolled mechanical damping ratio $\zeta$ and natural frequency $\omega_n$ as $\alpha_m = (\omega^2-\omega_n^2)/(2 \zeta \omega \omega_n)$, so the $\alpha_m=0$ plant resonates ($\omega=\omega_n)$ passively at the wave frequency.

Amplitude limits make design more complicated. Substituting $(\mathcal{R,D},\alpha_m)^\mathrm{opt}$ into \eqref{eq:alpha} gives $\alpha=0$, but section \ref{sec:generic} established that high $|\alpha|$ is desirable to minimize the effect of amplitude limits on power. Specifically, if $\mathcal{L}=0$, maximum $|\alpha|$ requires $\alpha_m^2=1+\mathcal{D}/\mathcal{R}$ instead of $\alpha_m=0$. Meanwhile, the well-known Bode-Fano limit implies that for good broadband matching, $|\alpha|$ must instead be minimized, implying either $\alpha_m=0$ or $|\alpha_m|\rightarrow\infty$. Therefore, the three goals of maximizing matched power ($\min|\alpha_m|$), minimizing the effect of constraints ($\max|\alpha|$), and maximizing bandwidth ($\min|\alpha|$) conflict. For the best tradeoff, the plant $\alpha_m$ must maximize $\overline{P}_L$ across the wave spectrum, the subject of future control co-design work.

\section{Saturation Nonlinearities}\label{sec:nonlinear}
\begin{figure*}[!b]
    \centering
    \includegraphics[width=18.3cm]{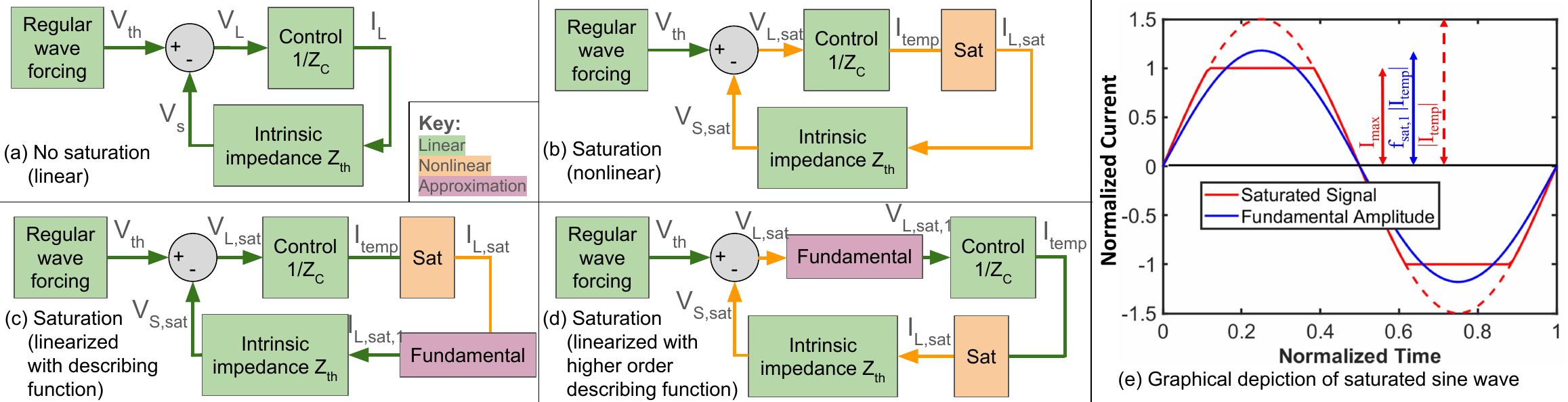}
    \caption{Block diagram illustration of describing functions used to linearize a wave energy saturation nonlinearity.}
    \label{fig:block-diagram-nonlinear}
\end{figure*}
\vspace{-5pt}Section \ref{sec:linear} considered a \textit{linear} impedance mismatch $Z_L = z Z_{th}^*$. In linear control, the current waveform is sinusoidal, so the fundamental amplitude equals the peak amplitude and never exceeds current limit $I_{max}$. 
\emph{Nonlinear} control allows non-sinusoidal current, increasing the fundamental. 

\vspace{-5pt}\subsection{Describing Functions}\label{sec:desc-fns}
\begin{figure}[htbp!]
    \centering\vspace{-5pt}
    \includegraphics[width=8cm]{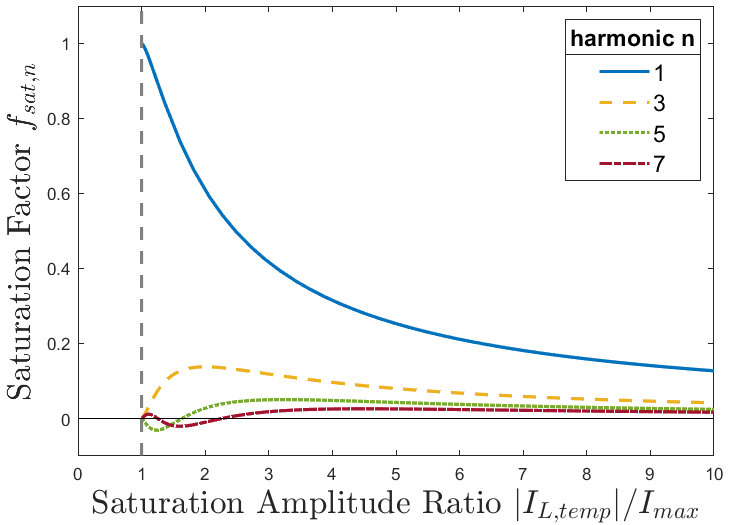}
    \vspace{-5pt}\caption{Harmonic ratios $f_{sat,n}$ as a function of saturation amplitude ratio $\mathcal{I}^{-1}=|I_{temp}|/I_{L,max}$, for n odd.}
    \label{fig:desc-fcn}
\end{figure}

\vspace{-5pt}Consider a nonlinear saturation control law of the form
\begin{equation}
    I_{L,sat}(t) = I_{max}~\textrm{sat}(I_{temp}(t)/I_{max})
\end{equation}
where $\textrm{sat}$ is the unit saturation function and $I_{temp}(t)$ is the time domain output of a linear impedance controller $I_{temp} = V_{L,sat}/Z_C$, stored temporarily in controller memory and not physically realized. 
Fig.~\ref{fig:block-diagram-nonlinear} illustrates this nonlinearity. Unlike the unsaturated case (a), the orange saturation blocks in cases (b)-(d) create harmonics. If $I_{temp}(t)$ is sinusoidal, $I_{L,sat}(t)$ is a saturated sine, shown in (e). Technically, $I_{temp}$ is nonsinusoidal because $I_{L,sat}$ harmonics propagate through linear blocks, shown with multiple orange lines in (b). However, the second-order low-pass plant dynamics $Z_{th}$ substantially reduce harmonics in downstream quantities $V_{S,sat}$, $V_{L,sat}$, and $I_{temp}$. Thus, (c)-(d) neglect harmonics of $I_{temp}$. Cases (c) and (d) differ in the location of the approximation: (c) depicts the sinusoidal-input describing function method from \cite{gelb_multiple-input_1968}, taking the fundamental of the saturated signal, and (d) is the higher order sinusoidal-input describing function from \cite{nuij_higher-order_2006}, preserving harmonics of $V_{L,sat}$. The higher order method is useful if the low-pass assumption is poor, which may be true for broadband WECs. The saturated-sine current signal is decomposed into a sum of harmonics:
\begin{equation}
    I_{L,sat}(t) \approx \sum_n |I_{L,sat,n}| \sin(n \omega t + \psi)
\end{equation}
where $\psi$ is the same phase as $I_{temp}(t)$, since saturation does not alter phase. The $n$th harmonic amplitude $|I_{L,sat,n}|$ is defined as $f_{sat,n} |I_{temp}|$. The saturation factor $f_{sat,n}$ is found via Fourier analysis:
\begin{equation}\label{eq:f-sat-desc-fcn}
    f_{sat,n} =
\begin{cases}
1 &  \mathcal{I}\geq 1,~n = 1\\
0 &  \mathcal{I}\geq 1, ~ n \neq 1 \\
\frac{2}{\pi} \left(\mathcal{I}\sqrt{1 - \mathcal{I}^2} + \theta \right) & \mathcal{I} <1, ~ n = 1 \\ 
0 & \mathcal{I}< 1,~ n = 2,4... \\
\frac{4}{\pi} \frac{n \sqrt{1 - \mathcal{I}^2}\sin\theta - \mathcal{I}\cos\theta}{n (n^2 - 1)}  &\mathcal{I}< 1 ,~ n = 3,5...
\end{cases}
\end{equation}
where $\mathcal{I}=I_{max}/|I_{temp}|$ and $\theta = n \sin^{-1}\mathcal{I} $. Fig.~\ref{fig:desc-fcn} depicts the first 7 harmonics. As the signal saturates, the fundamental decreases and higher harmonics emerge.

At each harmonic frequency $n$, the saturation block is approximated by gain $f_{sat,n}$, so the equivalent load impedance is $Z_L = Z_C / f_{sat,n}$ rather than the unsaturated $Z_L = Z_C$. This makes the nondimensional impedance for the $n$th harmonic $z_n = Z_C  / (f_{sat,n} Z_{th}^*)$. 
Now the response is the sum of linear responses at different frequencies:
\begin{equation}
    \overline{P}_L = \sum_{n} \overline{P}_{L,n}(z_n)
\end{equation}
In classic describing functions, only the fundamental current is used: $I_{L,sat} \approx I_{L,sat,1}$. This system contains only one harmonic, allowing the use of the tools from section \ref{sec:linear}, though it is still only quasi-linear because $f_{sat,1}$ depends on amplitude. The fundamental of the saturated current is a factor $f_{sat,1}/\mathcal{I}$ above the peak, thus nonlinear control allows higher current ratios for the same current limit. The resulting power increase is found from Fig.~\ref{fig:pareto} by moving on the x-axis a factor of $f_{sat,1}/\mathcal{I}$ higher than a given starting point. If $|I_{temp}|\gg I_{max}$, then $\mathcal{I}\rightarrow0$ and $f_{sat,1}/\mathcal{I}\rightarrow4/\pi$, yielding a square wave with fundamental $4/\pi$ higher than is possible in linear control.
To compute $\mathcal{I}$ otherwise, the dynamics are used to find complex-valued $I_{temp}$:
\begin{equation}\label{eq:f-sat-i-defn}
    \frac{I_{L,sat,1}}{I_{temp}} = \frac{V_{th}-V_{L,sat}}{I_{temp}Z_{th}} = \frac{V_{th}-Z_C I_{temp}}{I_{temp}Z_{th}}
\end{equation}
Controller impedance $Z_C$ is still undecided. A procedure aiming to maximize power might use equations \eqref{eq:ratios} and \eqref{eq:f-sat-desc-fcn} to select $Z_C$ so fundamental $|I_{L,sat,1}|$ approaches its unconstrained value $|I_L^m|$. However, invoking Pontryagin's principle avoids this effort. \cite{zou_optimal_2017} show analytically that steady-state optimal nonlinear control of a force-limited WEC is simply unconstrained optimal control with saturation, so $Z_C=Z_{th}^*$. \eqref{eq:f-sat-i-defn} is then solved using amplitude condition $f_{sat,1} = |I_{L,sat,1}/I_{temp}|$ and phase condition $\Im(I_{L,sat,1}/I_{temp}) = 0$. Combining with \eqref{eq:f-sat-desc-fcn} and \eqref{eq:thevenin} gives a transcendental equation. If an analytical solution is desired, arcsine can be algebraically approximated.

\subsection{Limitations and Extensions}\label{sec:limitations}
The sinusoidal-input describing function assumes a saturated-sine current waveform. This is reasonable for WECs (a) in regular waves, (b) with low-pass dynamics $|Z_{th}(n\omega)| \ll |Z_{th}(\omega)|$ for $n\geq3$. Critically, (a) fails in a broadband ocean environment. Accuracy in irregular waves remains to be tested. \cite{gelb_multiple-input_1968} suggest that sinusoidal-input describing functions still apply for some non-sinusoidal excitation, but \cite{rijlaarsdam_comparative_2017} show that systems with multiple sinusoids as input require other methods. Even for regular waves, the filter assumption (b) could be violated for broadband WECs, such as small WECs, or for narrow-band WECs analyzed far below the resonant frequency. This should be assessed with representative frequency-dependent hydrodynamics. Meanwhile, adding multiple constraints is unproblematic since arbitrarily many amplitudes may be analyzed/limited in the linear system, as long as all nonlinearity is filtered by subsequent low-pass dynamics. However, bump-stop position saturation is one problematic example where the nonlinearity occurs after the dynamics. Future work includes an error analysis compared to fully nonlinear simulations.

\section{Conclusion}
This work presented an analytical method to handle WEC generator force (current) constraints. First, linear theory was reviewed, and key relationships for an impedance-mismatched Thévenin-equivalent circuit were visualized on Smith charts and Pareto fronts. Then, specific wave energy dynamics were introduced, highlighting the implications of nondimensional resistance, damping, and mechanical impedance on power and sensitivity to amplitude limits. Next, describing function theory for the saturation nonlinearity was introduced, using the plant's low-pass nature to justify neglecting higher harmonics.

A variety of future work is possible. The effect of irregular waves and frequency-dependent dynamics on describing function accuracy should be investigated. The linear framework for voltage and current limits could easily extend to other constraints like position, velocity, phase voltage, or apparent power, and multiple degrees of freedom could be considered. When combined with cost estimates, this offers a powerful strategy for techno-economic tradeoff analysis and design optimization. All in all, while analytical approximations do not take the place of nonlinear numerical optimization, they provide the intuition and computational speed that may unlock improved WEC designs with substantial climate impact.

The code for this work is available open-source at \url{https://github.com/symbiotic-engineering/IFAC_CAMS_2024/}.

\bibliography{references}


\end{document}